\documentstyle[12pt]{article}
\newcommand{\beq}{ \begin{equation}}
\newcommand{\eeq}{\end{equation}}
\newcommand{\ba}{\begin{array}}
\newcommand{\ea}{\end{array}}
\newcommand{\beqa}{\begin{eqnarray}}
\newcommand{\eeqa}{\end{eqnarray}}

\newcommand{\rf}[1]{(\ref{#1})}
\def\Sum{\displaystyle\sum}

\font\twelve=cmbx10 at 15pt
\font\ten=cmbx10 at 12pt

\begin{document}

\begin{titlepage}

\begin{center}

\renewcommand{\thefootnote}{\fnsymbol{footnote}}

{\ten Centre de Physique Th\'eorique\footnote{
Unit\'e Propre de Recherche 7061} - CNRS - Luminy, Case 907}
{\ten F-13288 Marseille Cedex 9 - France }

\vspace{1,5 cm}

{\twelve EXISTENCE OF LONG-RANGE ORDER \\
IN QUASI-TWO-DIMENSIONAL HUBBARD MODEL}

\vspace{0.3 cm}

\setcounter{footnote}{0}
\renewcommand{\thefootnote}{\arabic{footnote}}

{\bf A. BELKASRI\footnote{and Universit\'e d'Aix-Marseille II}
 and J.L. RICHARD}

\vspace{3 cm}

{\bf Abstract}

\end{center}

In recent work of Monthoux and Pines~\cite{pines93} and also in Rice
et {\em al.}'s work~\cite{riceal}, quasi-averages like
$\langle c_{k \uparrow} c_{- k \downarrow} \rangle$ were considered
even in the case of a dimension less or equal two. But it is well
known from the old work of Hohenberg~\cite{hohen} that these
quasi-averages are zero at
$T \not= 0$ in case of 1 and 2 dimensions. In this communication we
apply the result of Hohenberg to the Hubbard model and prove that in
the case of quasi-two-dimension, the inequality of Bogoliubov is not
in contradiction with having
$\langle c_{k \uparrow} c_{- k \downarrow} \rangle \not= 0$ (at
$T \not= 0$) even for a system of three layers.

\vspace{3 cm}


\bigskip

\noindent July 1994

\noindent CPT-94/P.3025

\bigskip

\noindent anonymous ftp or gopher: cpt.univ-mrs.fr

\end{titlepage}

\noindent The models which are generaly used to describe the
electronic and magnetic properties of oxides
$H T_c$ superconductors are based on the Hubbard model. The
$H T_c$ oxides present a layered structure and copper-oxide planes
are the main structural elements of these materials. Therefore the
relevant properties of these oxides refer to a two dimensional
systems.

We know form the  work of Hohenberg~\cite{hohen} that a two
dimensional Fermi (or Bose) system can not exhibit a Long-Range
Order (LRO) for
$T \not= 0$. Nevertheless recently many
authors~\cite{pines93,riceal,hotta} considered a LRO for models
derived from the Hubbard model even in two-dimensions.

In BCS theory superconductivity is associated with anomalous
averages
$\langle c_{k \uparrow} c_{- k \downarrow} \rangle$ which are non
zero by vertue of a broken symmetry (the conservation of particle
number). Indeed, the key feature of BCS theory is Cooper-pair
condensation. The pair of states
$( k \uparrow,\ - k \downarrow )$ is occupied coherently. The
Cooper-pair amplitude,
$\langle c_{k \uparrow} c_{- k \downarrow} \rangle$, which is zero in
the normal state, becomes finite below
$T_c$.

\noindent In what follows, we repeat the proof of
Hohenberg~\cite{hohen} for the case of Hubbard model and show that in
quasi-two-dimensional case a LRO may exist.

\section{LRO in Hubbard model}

\subsection{ Hubbard model of fermions}

In this section we will examine the ordering for fermions in the
frame work of a quasi-two-dimensional Hubbard model. Let us consider
the following Hamiltonian
$H = H_t + H_U$, with
\beq
H_t = \sum_{ij} t_{ij} c^+_{i \sigma} c_{j \sigma}
\label{p4.1}
\eeq
and
\beq
H_U = \sum_{i,j \, ; \, \sigma, \sigma'} U_{ij} n_{i \sigma}
n_{j \sigma'}
\label{p4.2}
\eeq

\noindent where
$c_{i \sigma},\ c^+_{i \sigma}$ are the destruction and creation
operators, respectively, and
$n_{i \sigma} = c^+_{i \sigma} c_{i \sigma}$ with
$\sigma$ is the spin index. The hopping amplitude
$t_{ij}$ is nonzero only for
$(i,\ j)$ nearest-neighbors, and we have
\beq
t_{ij} = \left\{
\normalbaselineskip=18pt
\matrix{
- t & \mbox{for} & i - j = \pm e_x \ \, \mbox{or} \: \pm e_y \cr
- t' & \mbox{for} & i - j = \pm e_z \hfill \cr
0 & \;\;\;\mbox{otherwise} & \cr
}
\right.
\label{p4.3}
\eeq

\noindent with
$( e_x,\ e_y,\ e_z )$ the unit vectors for the three directions of
the lattice. The amplitude of the interaction
$U_{ij}$ is effective only in directions
$( e_x,\ e_y )$.

\noindent Let us define the charge density operator
\beq
\rho (i) = \sum_{\sigma} c^+_{i \sigma} c_{i \sigma}
\label{p4.4}
\eeq

\noindent Its $q^{\mbox{\footnotesize th}}$ Fourier component is
given by
\beq
\rho_q = \sum_{k, \sigma} c^+_{k, \sigma} c_{k + q, \sigma}
\label{p4.5}
\eeq

\noindent Since we will examine the existence or not of the
quasi-averages
$\langle c_{k \uparrow} c_{- k \downarrow} \rangle$ in one or more
regions of
$k$, we introduce the following order parameter
\beq
\Delta \equiv {1\over N} \sum_k S (k)
\langle c_{k \uparrow} c_{- k \downarrow} \rangle
\label{p4.6}
\eeq

\noindent where the ``smearing function''
$S (k)$ is arbitrary (a Gaussian, for instance) but has the
properties
\beq
S (0) = 1, \quad {1\over N} \sum_k S (k) < \infty
\label{p4.7}
\eeq

\noindent where
$N$ is the volume.

\noindent Now we will use the Bogoliubov inequality~\cite{hohen}
which is
\beq
\langle \left\{ A,\ A^+ \right\} \rangle \langle \left[ [ C,\ H ],\
C^+
\right] \rangle
\ge 2 k_B T | \langle [ C,\ A ] \rangle |^2
\label{p4.8}
\eeq

\noindent where
$\{ .,. \}$ is an anti-commutator,
$ [ .,. ] $ is a commutator,
$H$ is any Hamilonian,
$C,\ A$ are operators and finally
$\langle \dots \rangle$ indicates a statistical average on a grand
canonical ensemble
\beq
\langle X \rangle = Tr \left( Xe^{- \beta H}
\right) / Tr e^{- \beta H}
\eeq

\noindent straightforward algebraic calculation leads to
\beq
\left[ [ \rho_q,\ H_t ],\ \rho^+_q \right] =
- 2 \Sum_{i, j, \sigma} t_{ij} [ 1 - \cos q (i - j) ]
c^+_{i \sigma} c_{i \sigma}
\label{p4.9}
\eeq

\noindent In Fourier transform we obtain
\beq
\ba{ll}
\left[ [ \rho_q,\ H_t ],\ \rho^+_q \right] =
& 4 t \Sum_{k \sigma}
\left\{ ( 1 - \cos q_x ) \cos k_x + \right. \\
\vphantom{\Sum_{k \sigma}} & \left. \!  + ( 1 - \cos q_y ) \cos k_y +
\eta ( 1 - \cos q_z ) \cos k_z \right\} n_{k \sigma}

\ea
\label{p4.10}
\eeq
with
$\eta = t' / t$. Since
$H_U$ commute with
$\rho_q$, it is easy to get the following inequalities
\beq
0 < \langle \left[ [ \rho_q,\ H ],\ \rho^+_q \right] \rangle \le 4 t
\left\{ ( 1 - \cos q_x )
  + ( 1 - \cos q_y ) +
\eta ( 1 - \cos q_z ) \right\}
\Sum_k \langle n_k \rangle
\label{p4.11}
\eeq

\noindent and we have
\beq
\Sum_k \langle n_k \rangle = n N
\eeq
where
$n$ is the density of particles.

\noindent If in Bogoliubov inequality~(Eq.~\rf{p4.8}), we set
\beq
C = \rho_q \quad \mbox{and} \quad
A = {1 \over \sqrt N} \Sum_k S (k) c_{k \uparrow}
c_{- k + q \downarrow}
\label{p4.12}
\eeq

\noindent we  get
\beq
\langle \left\{ A,\ A^+ \right\} \rangle
\ge {{2 k_B T} \over {4\;t\;n}}\;
{
| \Delta_+ \Omega (q) |^2 \over
\left\{ ( 1 - \cos q_x ) + ( 1 - \cos q_y )
+ \eta ( 1 - \cos q_z ) \right\}
}
\label{p4.13}
\eeq
with
$\Omega (q) = \sum_k S (k + q)
\langle c_{k \uparrow} c_{- k \downarrow}
\rangle / N$. We notice that
$\Omega (q) \to \Delta$ when
$q \to 0$.

\noindent In the other hand, we have
$$
\langle \left\{ A,\ A^+ \right\} \rangle
= F (q) + R (q)
$$
with
\beq
F (q) = {2 \over N} \Sum_{k, k'} S (k) S^{\star} (k')
\langle c_{k \uparrow} c_{- k + q \downarrow}
c^+_{- k' + q \downarrow} c^+_{k' \uparrow} \rangle
\label{p4.14}
\eeq
and
\beq
R (q) = - {1 \over N} \Sum_k | S (k+q) |^2
\left\{ \langle n_{k \uparrow} \rangle +
\langle n_{- k \downarrow} \rangle \right\}+{1 \over N} \Sum_k | S
(k) |^2
\label{p4.15}
\eeq
$R (q)$ is a regular function of
$q$.

\noindent It is easy to see that
\beq
\langle c^+_{k \uparrow} c^+_{- k + q \downarrow}
c_{- k' + q \downarrow} c_{k' \uparrow} \rangle \le 1
\label{p4.16}
\eeq
and we  deduce that
\beq
{1 \over N} \Sum_{q \not= 0} F (q) < \infty
\label{p4.17}
\eeq

\noindent where we have use Eq.~\rf{p4.7}. The inequality~\rf{p4.13}
becomes
\beq
{1 \over N} \Sum_{q \not= 0} F (q) \ge
{k_B T \over {2 n t}}
{1 \over N} \Sum_{q \not= 0} {
| \Delta_+ \Omega (q) |^2 \over
\left\{ ( 1 - \cos q_x ) + ( 1 - \cos q_y )
+ \eta ( 1 - \cos q_z ) \right\}
} -
 \kern-0,34cm - {1 \over N} \Sum_{q \not= 0} R (q)
\label{p4.18}
\eeq

\noindent Since
$\Omega (q)$ and
$R (q)$ are regular in
$q$, at small
$q$ Eq.~\rf{p4.18} becomes
\beq
{1 \over N} \Sum_{q \not= 0} F (q) \ge
{4 k_B T \Delta^2 \over {n t}}
{1 \over N} \Sum_{q \not= 0} {
1 \over
q^2_x + q^2_y + \eta q^2_z
} - \Sum_{q \not= 0} R (q)
\label{p4.19}
\eeq

\noindent If now, we take the infinite volume limit for a layered
system (with
$2 L + 1$ planes and with periodic boundary conditions), we will have
\beqa
\nonumber
& & {1 \over 2 L + 1}
\Sum^L_{{n_z = - L}\atop {n_z\neq 0}}
 {1 \over (2 \pi)^2}
\int d q_x d q_y F(q) \ge
 {4 k_B T \Delta^2 \over n t}\times\\
& & {1 \over 2 L + 1}\Sum^L_{{n_z = - L}\atop {n_z\neq 0}}{1 \over
(2 \pi)^2}
\int d q_x d q_y
 { 1 \over
q^2_x + q^2_y + \eta \left( {2 \pi n_z \over 2 L + 1} \right)^2} - \\
& & {1 \over 2 L + 1}\Sum^L_{{n_z = - L}\atop {n_z\neq 0}}{1
\over (2 \pi)^2}
\int d q_x d q_y R(q)
\label{p4.20}
\eeqa

It is clear that in strictly two dimensional Hubbard model
($L = 0$
and
$\eta = 0$) the righthandsome of Eq.~\rf{p4.20} diverges and this in
contradiction with Eq.~\rf{p4.17} unless
$\Delta = 0$ (of course for the case
$T \not= 0$). Since
$S (k)$ is arbitrary this means that we cannot have
$\langle c_{k \uparrow} c_{- k \downarrow} \rangle \not= 0$ for any
$k$.

In the other hand, for quasi-two dimensional case (even for 3
layers~:
$L = 1$), the Hohenberg proof~(Eq.~\rf{p4.20}) cannot exclude to
have
$\langle c_{k \uparrow} c_{- k \downarrow} \rangle \not= 0$.

Then if someone believes that high
$T_c$ superconductivity is related to average like
$\langle c_{k \uparrow} c_{- k \downarrow} \rangle$, he should keep
in mind that he is dealing with a quasi-two-dimensional system and
not a strictly two-dimensional one.

In the case of the reduced BCS Hamiltonian quasi-averages
$\langle c_{k \uparrow} c_{- k \downarrow} \rangle$ exist even in 1
and 2 dimensions, because the interaction in this Hamiltonian is non
local (it depends on quasi-particles velocity) and then violates the
$f$-sum rule (which expresses the particle number
conservation)~\cite{ander58,bogo61}. In practice an other term coming
{}from the interaction is added to the Eq.~\rf{p4.11}.

\subsection{ Hubbard model of bosons}

The analogous situation holds also for the case of bosonic systems.
The extention of our investigation of LRO to Hubbard model for bosons
is also motivated by $H T_c$ superconductivity . Especially
the question of the existence or not of a bose
condensation for
$H T_c$ oxides~\cite{micnas}.

\noindent Let us consider the following Hamiltonian
\beq
H = \Sum_{ij} t_{ij} b^+_i b_j + \Sum_{ij} U_{ij} N_i N_j
\label{p4.21}
\eeq

\noindent where
$\left( b_i,\ b^+_i \right)$ is a boson field,
$N_i = b^+_i b_i$ and
$t_{ij}$ is given by~Eq.~\rf{p4.3}.

\noindent We define the charge density as
\beq
\rho_i = b^+_i b_i
\eeq
calculation leads to
\beqa
\nonumber
\left[ [ \rho_q,\ H ],\ \rho^+_q \right] &=& 4 t \Sum_k \left\{
( 1 - \cos q_x ) \cos k_x +\right.\\
&&\left.  ( 1 - \cos q_y ) \cos k_y + \eta ( 1 - \cos q_z )
\cos k_z \right\} N_k
\eeqa

\noindent and the following inequality holds
\beq
\ba{ll}
0 \langle \left[ [ \rho_q,\ H ],\ \rho^+_q \right] \rangle & \le 4 t
N \left\{ ( 1 - \cos q_x ) + ( 1 - \cos q_z ) + \right. \\
& \left. \hphantom{ \le 4 t N }
+ \eta ( 1 - \cos q_z ) \right\}
\ea
\eeq

\noindent If one uses the Bogoliubov inequality~Eq.~\rf{p4.8} for
\beq
C = \rho_q \quad \mbox{and} \quad A= b^+_q
\eeq

\noindent he gets the following result
\beq
{1\over N} \Sum_{q \ne 0} \langle N_q \rangle \ge -{1\over 2} +
{ k_B T | < b^+_i > |^2 \over 4 t }\;
{1\over N} \Sum_{q \ne 0}
{ 1 \over
( 1 - \cos q_x ) + ( 1 - \cos q_y ) + \eta ( 1 - \cos q_z ) }
\label{p4.22}
\eeq
Clearly we obtain the same result as for the Fermi systems. In
strictly two-dimen\-sio\-nal case
$\sum_{q \ne 0} < N_q > / N $ will be not bounded unless
$< b^+_i > = 0$, which implies that there is no LRO at
$T \ne 0$.
In quasi-two-dimensional case the inequality~\rf{p4.22} is not
in contradiction with having
$< b^+_i > \ne 0$ even at $T \ne 0$.

Finally we have proved that for the Hubbard model (of fermions or
bosons), the inequality of Bogoliubov which excludes to have  LRO in
 the two-dimensional case at
$T \ne 0$, does not exclude to have LRO for quasi-two-dimensional
case even for a finite number of  layers. We believe that writing
Eliashberg equations for models devoted to describe
$H T_c$ superconductivity will be meaningful only if we consider
quasi-two-dimensional systems.

\end{document}